\documentclass[5p]{elsarticle}

\usepackage{amsmath,amsthm,amssymb,amsfonts}
\usepackage[colorlinks,urlcolor=blue,linkcolor=blue,citecolor=blue,anchorcolor=blue]{hyperref}
\usepackage{lineno}
\modulolinenumbers[5]
\biboptions{sort&compress}

\journal{Science Bulletin}

\begin{document}
	
	\begin{frontmatter}
		
		\title{Observation of $\pi$ solitons in oscillating waveguide arrays}
		
		%%% Group authors per affiliation:
		\author[ad1,ad2]{Antonina A. Arkhipova\corref{equal}}
		\author[ad3]{Yiqi Zhang\corref{equal}}
		\cortext[equal]{These authors contributed equally to this work.}
		\author[ad1]{Yaroslav V. Kartashov\corref{cor}}
		\cortext[cor]{Corresponding author}
		\ead{yaroslav.kartashov@icfo.eu}
		\author[ad1,ad4]{Sergei A. Zhuravitskii}
		\author[ad1,ad4]{NNikolay N. Skryabin}
		\author[ad4]{Ivan V. Dyakonov}
		\author[ad1,ad4]{Alexander A. Kalinkin}
		\author[ad4]{Sergei P. Kulik}
		\author[ad1]{Victor O. Kompanets}
		\author[ad1]{Sergey V. Chekalin}
		\author[ad1,ad2]{Victor N. Zadkov}
		\address[ad1]{Institute of Spectroscopy, Russian Academy of Sciences, Troitsk, Moscow, 108840, Russia}
		\address[ad2]{Faculty of Physics, Higher School of Economics, Moscow, 105066, Russia}
		\address[ad3]{Key Laboratory for Physical Electronics and Devices of the Ministry of Education \& Shaanxi Key Lab of Information Photonic Technique, School of Electronic and Information Engineering, Xi'an Jiaotong University, Xi'an, 710049, China}
		\address[ad4]{Quantum Technology Centre, Faculty of Physics, M. V. Lomonosov Moscow State University, Moscow, 119991, Russia}

		\begin{abstract}
			Floquet systems with periodically varying in time parameters enable realization of unconventional topological phases that do not exist in static systems with constant parameters and that are frequently accompanied by appearance of novel types of the topological states. Among such Floquet systems are the Su-Schrieffer-Heeger lattices with periodically-modulated couplings that can support at their edges anomalous $\pi$ modes of topological origin despite the fact that the lattice spends only half of the evolution period in topologically nontrivial phase, while during other half-period it is topologically trivial. Here, using Su-Schrieffer-Heeger arrays composed from periodically oscillating waveguides inscribed in transparent nonlinear optical medium, we report experimental observation of photonic anomalous $\pi$ modes residing at the edge or in the corner of the one- or two-dimensional arrays, respectively, and demonstrate a new class of topological $\pi$ solitons bifurcating from such modes in the topological gap of the Floquet spectrum at high powers. $\pi$ solitons reported here are strongly oscillating nonlinear Floquet states exactly reproducing their profiles after each longitudinal period of the structure. They can be dynamically stable in both one- and two-dimensional oscillating waveguide arrays, the latter ones representing the first realization of the Floquet photonic higher-order topological insulator, while localization properties of such $\pi$ solitons are determined by their power.
		\end{abstract}
		
		
	\end{frontmatter}
	
	%\linenumbers
	
	\section{Introduction}
	Photonic topological insulators~\cite{lu.np.8.821.2014, ozawa.rmp.91.015006.2019} are unique materials hosting localized topologically protected states at their edges by analogy with edge modes in electronic topological insulators, first predicted in solid-state physics~\cite{hasan.rmp.82.3045.2010, qi.rmp.83.1057.2011}. Various mechanisms of formation of the photonic topological edge states were discovered, most of which are associated with breakup of certain symmetries of the underlying system possessing specific degeneracies in the linear spectrum. The most representative feature of topological edge states is their remarkable robustness with respect to deformations of the structure, disorder, and their persistence for different geometries of the edge between topologically distinct materials. Their formation and robustness has been predicted and demonstrated for various photonic systems with broken time-reversal symmetry, for valley-Hall systems with broken inversion symmetry, and in higher-order topological insulators~\cite{haldane.prl.100.013904.2008, wang.nature.461.772.2009, hafezi.nphys.7.907.2011, lindner.nphys.7.490.2011, rechtsman.nature.496.196.2013, khanikaev.nmat.12.233.2013, hafezi.np.7.1001.2013, mittal.prl.113.087403.2014, wu.prl.114.223901.2015, gao.nc.7.11619.2016, klembt.nature.562.552.2018, noh.prl.120.063902.2018, yang.nature.565.622.2019, lustig.nature.609.931.2022, pyrialakos.nm.21.634.2022, xie.natrevphys.3.520.2021}. Particularly nontrivial situation is realized when topological phase is induced by periodic modulations of system parameters in the evolution variable~\cite{garanovich.pr.518.1.2012}, for example in the direction of light propagation. Characterization of such driven Floquet systems usually requires special topological invariants, as shown in~\cite{kitagawa.prb.82.235114.2010, rudner.prx.3.031005.2013, rudner.nrp.2.229.2020}. Among the most intriguing manifestations of topological effects in Floquet systems is the formation of unidirectional edge states as proposed in~\cite{lindner.nphys.7.490.2011} and observed at optical frequencies in~\cite{rechtsman.nature.496.196.2013}, observation of anomalous topological states~\cite{maczewsky.nc.8.13756.2017, mukherjee.nc.8.13918.2017, leykam.prl.117.013902.2016}, and of so-called anomalous $\pi$ modes associated with nonzero $\pi$ gap invariant and studied in~\cite{asboth.prb.90.125143.2014, dallago.pra.92.023624.2015, fruchart.prb.93.115429.2016, zhang.acs.4.2250.2017, petracek.pra.101.033805.2020}. Recent surge of interest to topological pumping in near-solitonic regime should be mentioned too~\cite{jurgensen.nature.596.63.2021, fu.prl.128.154101.2022, fu.prl.129.183901.2022}.
	
	$\pi$ modes are unique topological states that may appear in a quasi-energy spectrum of the Floquet system that due to modulation of its parameters spends half of the evolution period in ``instantaneous'' nontopological phase, while on the other half of the period it is topologically nontrivial.
	In tight-binding models describing Floquet systems, $\pi$ modes usually appear at the edges of the ``longitudinal'' Brillouin zone with quasi-energies equal to $\pm \pi/T$ (where $T$ is period of the Floquet system), in contrast to conventional ``zero-energy'' edge states in static topological systems, hence the notion of $\pi$ modes that we also use in this work for convenience. So far, the photonic $\pi$ modes have been observed only in linear regime in one-dimensional modulated Su-Schrieffer-Heeger (SSH) arrays in a microwave range~\cite{cheng.prl.122.173901.2019} and at optical frequencies in non-Hermitian or plasmonic SSH arrays~\cite{wu.prr.3.023211.2021, song.lpr.15.2000584.2021, sidorenko.prr.4.033184.2022} with high refractive index contrast, where, however, considerable losses limit propagation distances to hundreds of micrometers. $\pi$ modes may have applications in the design of systems supporting high-quality cavity modes~\cite{ota.optica.6.786.2019, xie.lpr.14.1900425.2020}, for realization of low-threshold lasers~\cite{zhang.light.9.109.2020, kim.nc.11.5758.2020, zhong.apl.6.040802.2021, ota.nano.9.547.2020}, in strongly correlated electron–photon systems~\cite{bloch.nature.606.41.2022}, and other areas. They have been also encountered beyond the realm of optics, e.g., in acoustics~\cite{cheng.prl.129.254301.2022, zhu.nc.13.11.2022}. Nevertheless, to date the photonic $\pi$ modes remain unobserved in higher-dimensional conservative systems and their nonlinear analogs were never reported experimentally. At the same time, photonic Floquet systems offer unique testbed for the exploration of nonlinear effects in specifically designed low-loss topological guiding structures, where observation of so far elusive class of $\pi$ solitons is possible.
	
	It should also be mentioned that nonlinearity is playing an increasingly important role in all-optical control of topological systems, see recent reviews~\cite{smirnova.apr.7.021306.2020, ozawa.rmp.91.015006.2019}. In particular, nonlinearity may stimulate modulational instability of the nonlinear edge states~\cite{lumer.pra.94.021801.2016, leykam.prl.117.143901.2016, kartashov.optica.3.1228.2016}, it leads to rich bistability effects for edge states in pumped dissipative resonator structures~\cite{kartashov.prl.119.253904.2017, mandal.acs.10.147.2023, pernet.nphys.18.678.2022}, it may cause power-controlled topological transitions~\cite{maczewsky.science.370.701.2020}, and enables the formation of topological solitons both in the bulk of the insulator ~\cite{lumer.prl.111.243905.2013, mukherjee.science.368.856.2020} and at its edges~\cite{leykam.prl.117.143901.2016, ablowitz.pra.96.043868.2017, gulevich.sr.7.1780.2017, li.prb.97.081103.2018, smirnova.lpr.13.1900223.2019, zhang.prl.123.254103.2019, ivanov.acs.7.735.2020, zhong.ap.3.056001.2021, smirnova.prr.3.043027.2021, mukherjee.prx.11.041057.2021, xia.light.9.147.2020, guo.ol.45.6466.2020, kartashov.prl.128.093901.2022}. The important property of such solitons is that they remain localized due to nonlinearity, and at the same time they inherit topological protection from linear edge modes, from which they usually bifurcate. Corner solitons in higher-order topological insulators have been reported too~\cite{zangeneh.prl.123.053902.2019, kirsch.np.17.995.2021, hu.light.10.164.2021}. Very recently it was theoretically predicted~\cite{zhong.pra.107.L021502.2023} that Floquet topological systems may support a new class of topological solitons, qualitatively different from previously observed unidirectional states \cite{mukherjee.prx.11.041057.2021} -- namely, $\pi$ soliton -- that represents dynamically oscillating nonlinear Floquet state with a quasi-energy in the topological bandgap that exactly reproduces its intensity distribution after each longitudinal period of the structure. Even strongly localized $\pi$ solitons are practically free from radiative losses that usually restrict propagation distances for unidirectional edge solitons in Floquet waveguiding systems~\cite{ivanov.acs.7.735.2020, mukherjee.prx.11.041057.2021}.
	
	In this work, we report on the experimental observation of $\pi$ solitons in one- and two-dimensional Floquet waveguide arrays, where nontrivial topological properties arise due to $z$-periodic oscillations of waveguide centers in each unit cell of the structure. The arrays considered here are inscribed in a transparent nonlinear dielectric medium (fused silica) using the technique of direct femtosecond laser writing~\cite{tan.ap.3.024002.2021, li.ap.4.024002.2022, lin.us.2021.9783514.2021} and represent SSH-like structures, which, however, are not static, but spend half of the $z$-period in ``instantaneous'' topological phase, while during other half of the period they are ``instantaneously'' non-topological, as defined by periodically varying intra- and inter-cell coupling strengths. In two dimensions such arrays represent the realization of the photonic Floquet higher-order insulator. Floquet spectrum of such arrays is characterized by the presence of in-gap topological $\pi$ modes, from which robust $\pi$ solitons can bifurcate in the nonlinear regime. We observe such solitons using single-site excitations, study their periodic evolution with distance, and dependence of their localization properties on the amplitude of the waveguide oscillations and power.
		
	\begin{figure*}[htpb]
		\centering
		\includegraphics[width=0.8\textwidth]{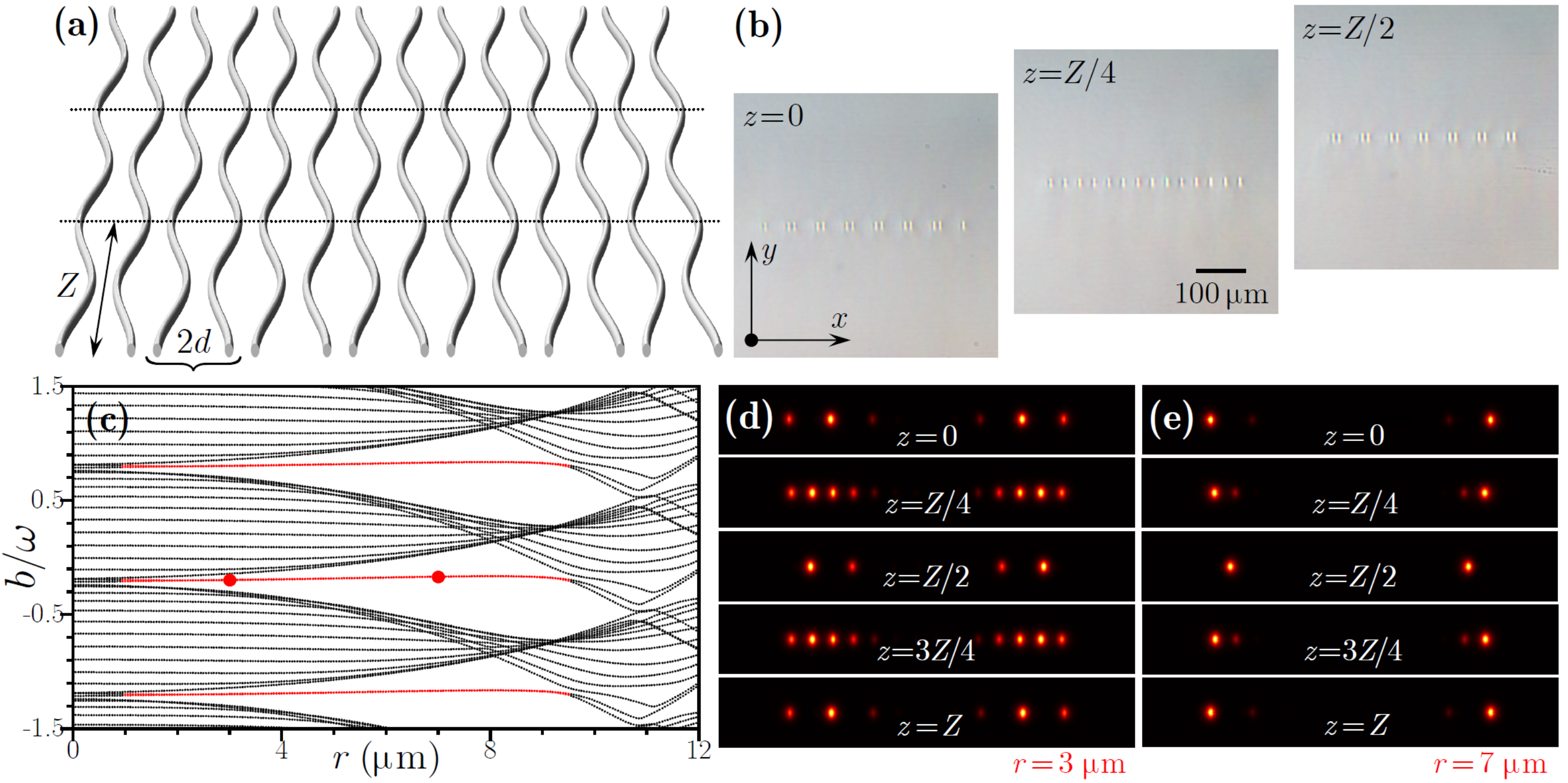}
		\caption{(Color online) (a) Schematic image of the 1D oscillating waveguide array (three longitudinal $z$-periods) containing 7 unit cells. (b) Microphotographs of the fs‐laser written oscillating waveguide array at different distances in topological phase $(z=0)$ , uniform phase $(z=Z/4)$, and trivial phase $(z=Z/2)$. (c) Quasi‐propagation constants of the Floquet modes of the oscillating array versus amplitude of the waveguide oscillations $r$ within three longitudinal Brillouin zones. (d) and (e) show intensity distributions of the linear $\pi$ modes at different distances $z$ for two selected oscillation amplitudes $r$ corresponding to the red dots in (c). In all cases $Z=33 \,\textrm{mm}$.}
		\label{fig1}
	\end{figure*}
	
	\section{Results and discussions}
	We consider paraxial propagation of a light beam along the $z$ axis of the medium with focusing cubic nonlinearity and shallow transverse modulation of the refractive index that can be described by the nonlinear Schr\"odinger-like equation for the dimensionless light field amplitude $\psi$:
	\begin{equation}\label{eq1}
		i \frac{\partial \psi}{\partial z} = -\frac{1}{2} \left( \frac{\partial^2}{\partial x^2} + \frac{\partial^2}{\partial y^2} \right) \psi
		-\mathcal{R}(x,y,z) \psi - |\psi|^{2} \psi.
	\end{equation}
	Here $x,y$ are the scaled transverse coordinates, $z$ is the propagation distance that plays in Eq.~(\ref{eq1}) the same role as time in the Schrödinger equation describing a quantum particle in a potential, and the function ${\mathcal R}(x,y,z)$ describes array of periodically oscillating waveguides.
	For details of normalization of Eq.~(\ref{eq1}) see Section S1 in the Supplementary materials .
	
	\subsection{1D $\pi$ solitons}
	
	First of all, for observation of 1D $\pi$ solitons we consider the SSH-like arrays of oscillating waveguides containing 7 unit cells. Refractive index distribution in such arrays can be described by the following function:
	\[
	{\mathcal R} = p \sum_{m} [e^{-(x^2_{1m}/a_x^2+y^2/a_y^2)} +  e^{-(x^2_{2m}/a_x^2+y^2/a_y^2)}],
	\]
	where $x_{1m} = x_m + d/2 + r\cos(\omega z)$ and $x_{2m} = x_m - d/2 - r\cos(\omega z)$ are the $x$-coordinates of the waveguide centers in each unit cell containing two waveguides, $\omega=2\pi/Z$ is the spatial frequency of oscillations of the waveguide centers, $Z$ is the longitudinal period of the array, $x_m=x-2md$, $m$ is the integer index of the cell, $r$ is the amplitude of the waveguide oscillations, which was varied from $1$ to $11\,\mu\textrm{m}$, $d=30\,\mu\textrm{m}$ is the spacing between waveguides at $r=0$ (i.e. unit cell size is $2d$), $a_x=2.5~\mu \textrm{m}$ and $a_y=7.5~\mu \textrm{m}$ are the widths of waveguides that are elliptical due to writing process, and $p$ is the array depth proportional to the refractive index contrast $\delta n$ in the structure (see Section S1 in the Supplementary materials ). Schematic illustration of such array is presented in Fig.~\ref{fig1}a. As one can see, the separation $d-2r\cos(\omega z)$ between two waveguides in the unit cell (intracell separation) of this structure varies dynamically, leading to periodic transformation between ``instantaneously'' topological (inter-cell coupling exceeds intracell one) and non-topological (inter-cell coupling is weaker than intra-cell one) SSH configurations. Microphotographs of such 1D fs-laser written arrays in fused silica at different distances within the sample are presented in Fig.~\ref{fig1}b. The array period $Z=33\,\textrm{mm}$ was selected such that our samples contained three full $z$-periods of this Floquet structure (see Section S2 in the Supplementary materials ).
	
	Nontrivial topological properties in this system arise due to longitudinal variations of the structure (oscillations of the waveguides). Its modes are the Floquet states $\psi=u(x,y,z) e^{ibz}$, where $b$ is a quasi-propagation constant [for first Brillouin zone $b\in [-\omega/2,+\omega/2)$], and $u(x,y,z)=u(x,y,z+Z)$ is the $Z$-periodic complex field that satisfy the equation
	\begin{equation}\label{eq2}
		bu = \frac{1}{2} \left( \frac{\partial^2}{\partial x^2} + \frac{\partial^2}{\partial y^2} \right) u + \mathcal{R}u + i \frac{\partial u}{\partial z} +|u|^2 u.
	\end{equation}
	Neglecting nonlinear term in Eq.~(\ref{eq2}) we calculate linear spectrum of 1D array using the method proposed in~\cite{leykam.prl.117.013902.2016} (see Section S5 in the Supplementary materials ). The transformation of linear spectrum with increase of the amplitude $r$ of waveguide oscillations is shown in Fig.~\ref{fig1}c. Quasi-propagation constant $b$ is defined modulo $\omega$ and in Fig.~\ref{fig1}c we show the spectrum within three longitudinal Brillouin zones. Gray lines correspond to the delocalized bulk modes, while the red lines correspond to the linear topological $\pi$ modes~\cite{asboth.prb.90.125143.2014, dallago.pra.92.023624.2015, fruchart.prb.93.115429.2016, zhang.acs.4.2250.2017, petracek.pra.101.033805.2020}. Notice that they emerge around the points, where Floquet replicas of the bands spectrally overlap, and longitudinal modulation hybridizes states at the band edges lifting their degeneracy and opening the topological gap. Because our structure is symmetric, $\pi$ modes appear near both edges of the array. Quasi-propagation constants of $\pi$ modes are located in the forbidden gap in the Floquet spectrum that guarantees absence of coupling with bulk states. Their localization near a given edge increases with increase of the gap width, cf Fig.~\ref{fig1}d and  e. Such modes show strong shape transformations within longitudinal period, but exactly reproduce their shape after each period $Z$. Remarkably, Fig.~\ref{fig1}e clearly shows that global intensity maximum of the $\pi$ mode is not always located in the edge waveguide. For instance, at $z=Z/2$, where the array is in instantaneous nontopological phase, the intensity maximum switches into next to edge waveguide, while at $z=Z$, exactly after one oscillation period, where structure returns into instantaneous topological configuration, the light also switches back to the edge waveguide. Already for $r = 7\,\mu\textrm{m}$ the $\pi$ mode contracts practically to single waveguide in some points within evolution period that enables its efficient excitation in the experiment [this determined our choice of the initial ``phase'' of the waveguide oscillations in Fig.~\ref{fig1}a; linear spectrum clearly does not depend on this phase]. Very similar results and Floquet spectrum were obtained also for arrays with odd number of waveguides (where one of the unit cells is incomplete), because even in this case due to waveguide oscillations both edges periodically pass through stages when truncation becomes topological or nontopological and therefore support $\pi$ modes.
	
	The appearance of topological $\pi$ modes in the spectrum of this Floquet system is associated with nonzero value of the $\pi$ gap invariant $w_{\pi}$ (see Section S4 in the Supplementary materials  for details of its calculation and literature~\cite{fruchart.prb.93.115429.2016,cheng.prl.122.173901.2019}). One can observe the formation of $\pi$ modes in the spectrum of truncated array when $w_{\pi}=1$ (for sufficiently large oscillation amplitudes $r$), while these modes are absent when $w_{\pi}=0$ (e.g., at $r\to 0$).
	
	Inspecting spectrum in Fig.~\ref{fig1}c one can see that while longitudinal modulation with frequency $\omega$ creates Floquet replicas of bands and localized $\pi$ modes, by itself it does not induce parametric resonances between localized and bulk modes, which nevertheless can occur, if additional weak modulation of optical potential at frequencies $2\omega$, $3\omega$, $\cdots$ is added that keeps $Z$-periodicity of array.
	
\begin{figure}[htpb]
		\centering
		\includegraphics[width=\columnwidth]{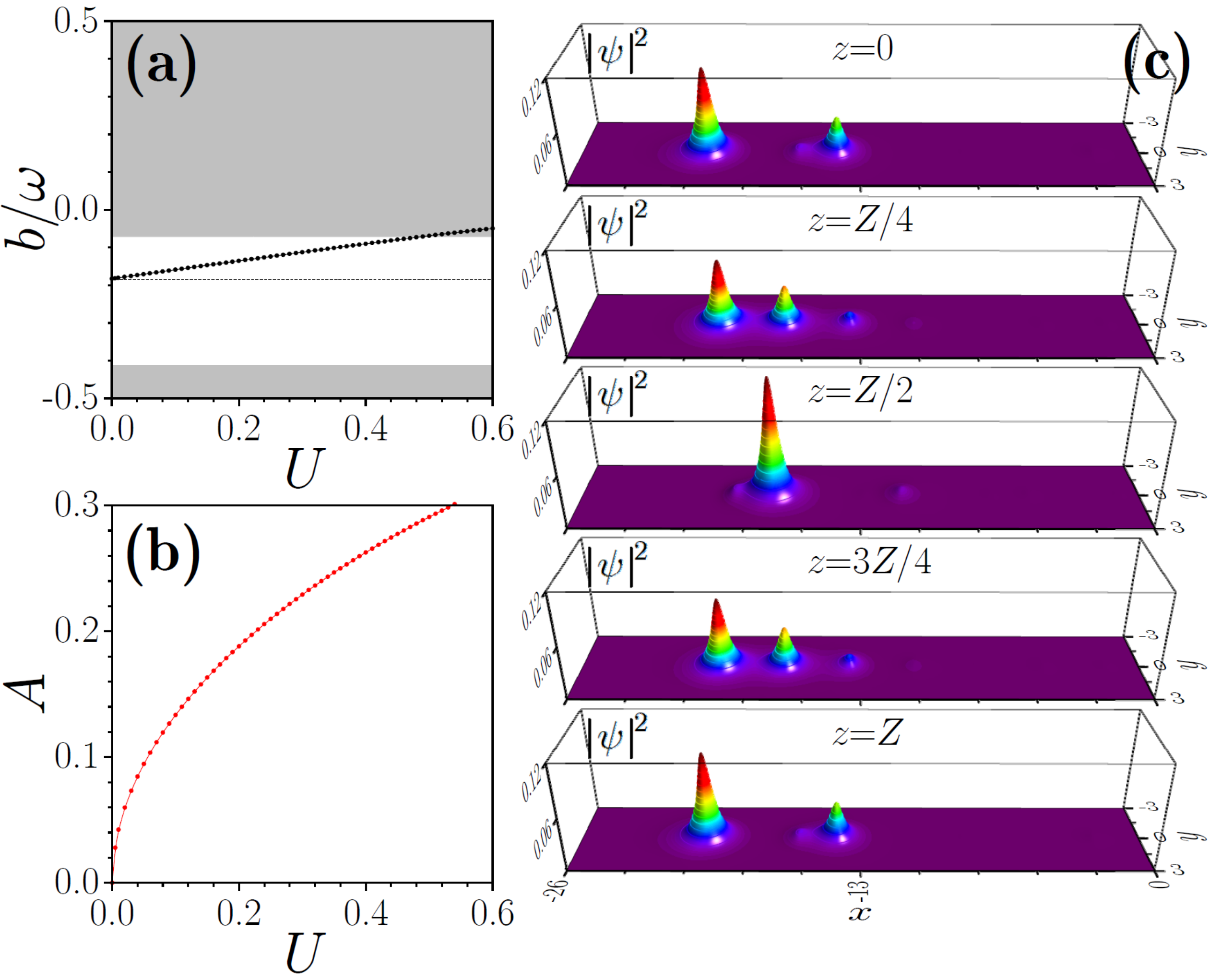}
		\caption{(Color online) Quasi‐propagation constant (a) and $z$‐averaged peak amplitude (b) of the $\pi$ soliton versus its power $U$. Gray region in (a) corresponds to the bulk band, while white region corresponds to the forbidden gap in Floquet spectrum. Horizontal dashed line shows quasi-propagation constant $b$ of linear $\pi$ mode. (c) Intensity distributions in the $\pi$ soliton with power $U =0.5$ at different distances within one oscillation period $Z$. The state is shown near left edge only. Here $r = 6\,\mu\textrm{m}$, $Z=33 ~\textrm{mm}$. The dependencies $b(U)$ and $A(U)$ for other values of $3\,\mu\textrm{m} \le r \le 9\,\mu\textrm{m}$ are qualitatively similar.}
		\label{fig2}
	\end{figure}

	The $\pi$ solitons are the topological nonlinear Floquet states bifurcating from the linear $\pi$ modes. To find their profiles we iteratively solve Eq.~(\ref{eq2}) with the last nonlinear term included (see Section S6 in the Supplementary materials ), by varying soliton power $U=\iint |\psi|^2dxdy$ and calculating for each $U$ corresponding $Z$-periodic soliton profile $u(x,y,z)$, quasi-propagation constant $b$, and averaged amplitude $A=Z^{-1}\int_z^{z+Z} \textrm{max}|\psi|dz$. The $\pi$ solitons bifurcate from linear $\pi$ modes, as evident from the representative $b(U)$ dependence in Fig.~\ref{fig2}a, where quasi-propagation constant of linear mode is shown by the dashed line. Their amplitude $A$ increases with the power (Fig.~\ref{fig2}b). Importantly, nonlinearity changes the location of $b$ inside the topological gap, gradually shifting it towards the bulk band (gray region). This is accompanied by changes in soliton localization in the $(x,y)$ plane (it may first increase and then decrease depending on the value of $r$), especially when $b$ shifts into the band, where coupling with the bulk modes occurs. Periodic transformation of soliton intensity distribution with $z$ is illustrated in Fig.~\ref{fig2}c. It should be stressed that for our parameters even for the amplitude of oscillations $r\sim9~\mu \textrm{m}$ the $\pi$ solitons obtained here are robust objects that practically do not radiate and survive over hundreds of $Z$ periods that is beneficial in comparison with the previously observed unidirectional edge solitons~\cite{mukherjee.prx.11.041057.2021}.
	
	We tested stability of such states by adding broadband small noise (typically, $5 \%$ in amplitude) into input field distributions and propagating such perturbed $\pi$ solitons over distances $\sim 500Z$ that allows to detect the presence of even very weak instabilities. Such stability analysis has shown that for our parameters and for amplitudes of oscillations $r>5~\mu\textrm{m}$ 1D solitons belonging to forbidden gap are stable, while they become unstable, when they shift into band. Notice that stabilization of such states that have powers well below power of Townes soliton in uniform cubic medium is consistent with arguments of Ref. \cite{sakaguchi.pre.89.032920.2014}.
	
	\begin{figure*}[htpb]
		\centering
		\includegraphics[width=0.99\textwidth]{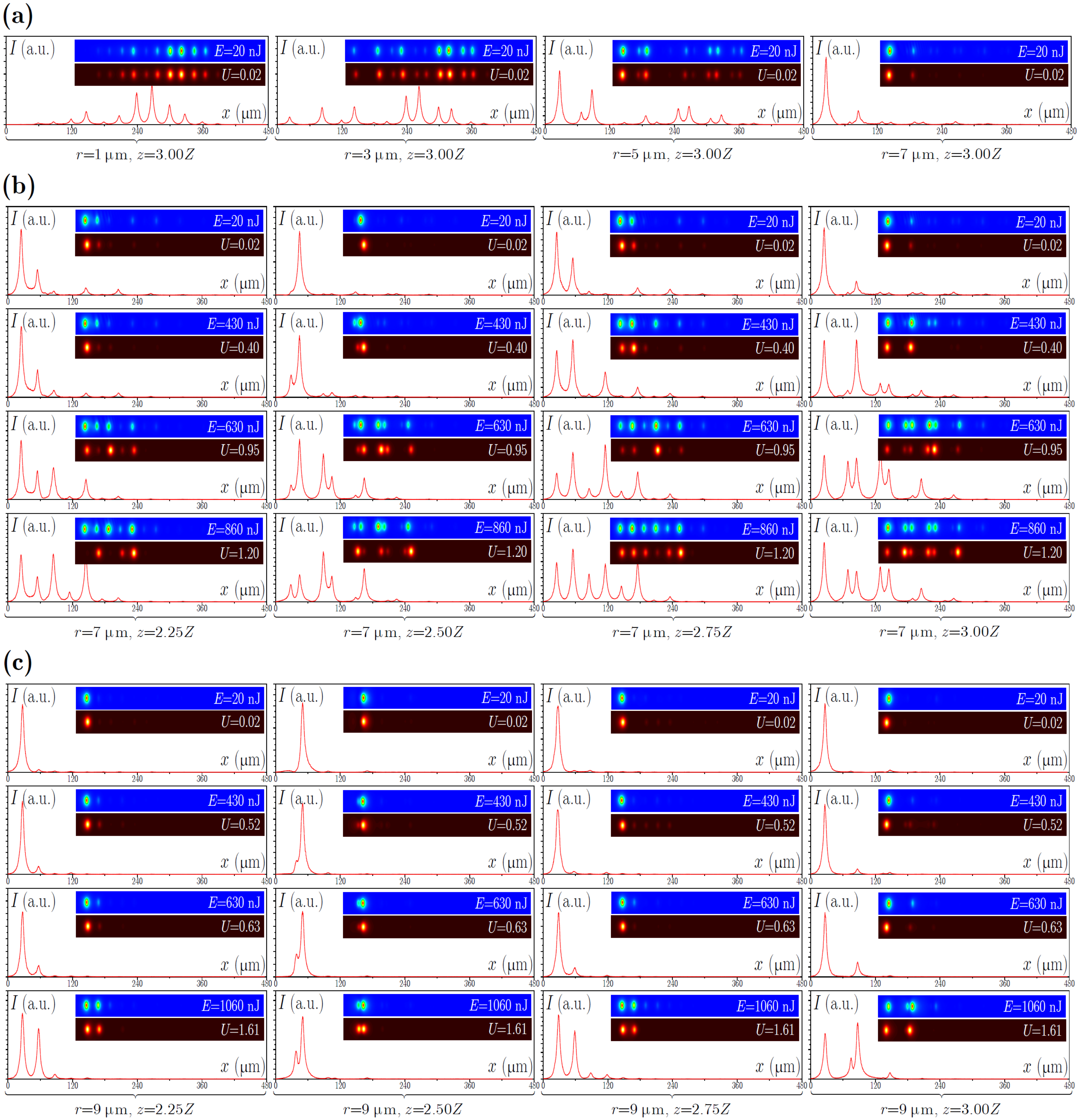}
		\caption{(Color online) (a) Formation of the $\pi$ modes with increase of the amplitude of waveguide oscillations $r$ in linear regime. The impact of nonlinearity on such states and formation of $\pi$ solitons is illustrated in (b) for $r=7\,\mu\textrm{m}$ and in (c) for $r=9\,\mu\textrm{m}$ at different propagation distances. In all panels red lines show experimental 1D intensity cross-sections at $y=0$, blue inset shows 2D experimental intensity distributions for a given pulse energy $E$, and black-red insets show corresponding theoretical 2D intensity distributions for different input powers $U$.
		}
		\label{fig3}
	\end{figure*}
	
	To demonstrate 1D $\pi$ solitons experimentally we inscribed in a fused silica sample the series of SSH-like arrays with the different amplitudes $r$ of waveguide oscillations ranging from $1$ to $11~\mu\textrm{m}$, with a step in $r$ of $2~\mu\textrm{m}$ using fs-laser writing technique (see Fig.~\ref{fig1}b with exemplary microphotographs of the array and Section S2 in the Supplementary materials  for the details of inscription). While full sample length contains three $Z$-periods of the structure, to demonstrate dynamics in the internal points of the last period, we additionally inscribed arrays with fractional lengths $2.25Z$, $2.50Z$, $2.75Z$ with the same parameters (see Section S2 in the Supplementary materials  for the details).
	
	\begin{figure*}[htpb]
		\centering
		\includegraphics[width=0.9\textwidth]{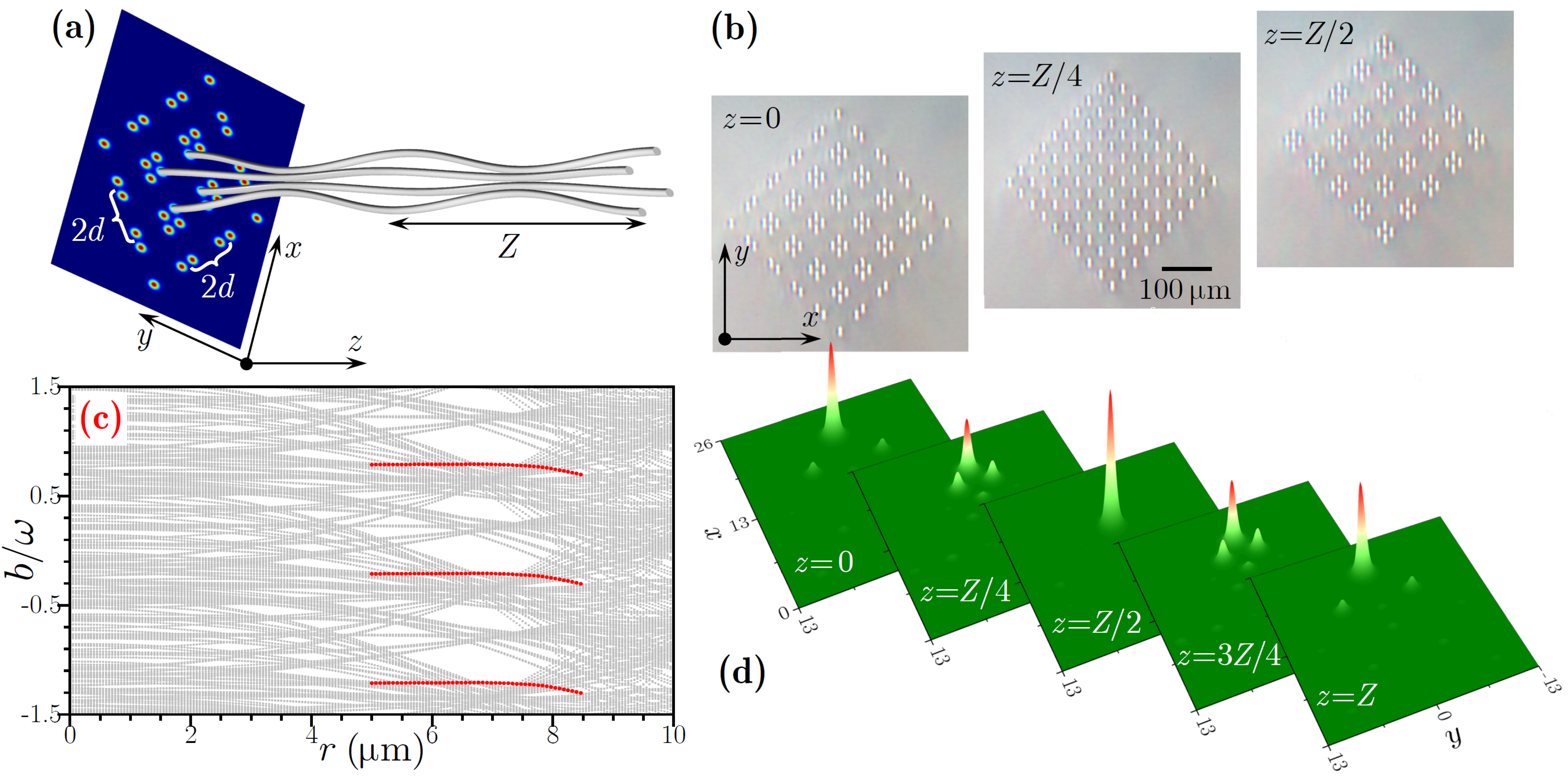}
		\caption{(Color online) (a) Schematic illustration of the 2D oscillating waveguide array (for illustrative purposes we show only $3 \times 3$ unit cells). (b) Microphotographs of laser‐written 2D array with $5 \times 5$ unit cells at different distances. (c) Quasi‐propagation constants of the Floquet modes of the 2D array with oscillating waveguides versus amplitude of waveguide oscillations $r$ within three longitudinal Brillouin zones. (d) Intensity distribution in linear 2D $\pi$ mode at different distances $z$ for $r = 6\,\mu\textrm{m}$. In all cases $Z=49.5\,\textrm{mm}$.}
		\label{fig4}
	\end{figure*}
	
	In experiments we excited the waveguide at the left edge using the fs-laser pulses of variable energy $E$ (for correspondence between pulse energy $E$ and input peak power in the waveguide see Section S3 in the Supplementary materials ). Output intensity cross-sections at $y=0$ (red lines) and 2D distributions (blue insets) are compared in Fig.~\ref{fig3} with the results of theoretical simulations of the single-site excitation with different input powers $U$ in the frames of Eq.~(\ref{eq1}) (black-red insets). In all cases theoretical results well agree with the experimental observations. In Fig.~\ref{fig3}a and top right image in Fig.~\ref{fig3}c we show how output beam localization progressively increases in linear regime (low pulse energies $E=20\,\textrm{nJ}$) with increase of the amplitude $r$  of the waveguide oscillations. Efficient excitation of well-localized linear $\pi$ modes is evident for $r\ge 7\,\mu\textrm{m}$, while for $r=5\,\mu\textrm{m}$ the excitation efficiency is lower and some fraction of power penetrates into the bulk of the array. First rows in Fig.~\ref{fig3}b and c illustrate that $\pi$ mode undergoes strong oscillations on one $Z$ period, main intensity maximum switches into second waveguide at $z=2.50Z$ (consistently with the dynamics of exact state in Fig.~\ref{fig1}e), but it returns to the edge one at $z=3.00Z$ illustrating periodic evolution.

	By increasing pulse energy we observe the formation of the 1D $\pi$ solitons. As mentioned above, they are in-gap topological states bifurcating from $\pi$ modes under the action of nonlinearity. Nonlinearity leads to soliton reshaping (in particular, for $r=7-9\,\mu\textrm{m}$ it slightly broadens with increase of $U$), but when its quasi-propagation constant shifts into the band, strong radiation into the bulk occurs. This is most clearly visible for $r=7\,\mu\textrm{m}$ (Fig.~\ref{fig3}b), where solitons were observed well-localized near the edge for the pulse energies $E<350\,\textrm{nJ}$, but radiating around $E\sim430\,\textrm{nJ}$ (notice that at this energy the level of radiation becomes visible only after three $Z$ periods). Further increase of the pulse energy leads to stronger radiation. At $r=9\,\mu\textrm{m}$ the range of the pulse energies, where formation of robust $\pi$ solitons is observed substantially increases (Fig.~\ref{fig3}c). Well localized $\pi$ solitons performing $Z$-periodic oscillations are observed for the pulse energies $E<900\,\textrm{nJ}$ (rows 1-3) and only at $E\sim1000\,\textrm{nJ}$ small radiation due to the coupling with the bulk modes appears (row 4). Simulations over much larger distances $(z>100Z)$ confirm robustness of such dynamically excited nonlinear Floquet states with in-gap quasi-propagation constants. It should be stressed that excitation of the waveguide in the bulk of the above arrays does not yield localization for considered pulse energies.
	
	\subsection{2D $\pi$ solitons}
	
	For observation of 2D $\pi$ solitons we utilize 2D generalization of the SSH array with oscillating waveguides. Unit cell of such an array (quadrimer) contains four waveguides, whose centers oscillate with period $Z$ along the diagonals of the unit cell. We consider sufficiently large structure containing $5\times5$ unit cells. Refractive index distribution in this Floquet structure is described by the function
	\[
	\begin{split}
		{\mathcal R}  =p \sum_{m,n} [& e^{-(x^2_{1m}/a_x^2+y^2_{1n}/a_y^2)} + e^{-(x^2_{2m}/a_x^2+y^2_{1n}/a_y^2)} + \\
		&e^{-(x^2_{1m}/a_x^2+y^2_{2n}/a_y^2)} + e^{-(x^2_{2m}/a_x^2+y^2_{2n}/a_y^2)}],
	\end{split}
	\]
	where  $x_{1m,2m} = x_m \pm d/2 \pm r\cos(\omega z)$ and ${y_{1n,2n} = y_n} \pm d/2 \pm r \cos(\omega z)$ are the coordinates of centers of four waveguides in the unit cell with $x_m=x-2md$ and $y_n=y-2nd$, $m,n$ are the integers. In 2D case the oscillation period was taken as $Z=49.5\,\textrm{mm}$, so that sample contained two full longitudinal array periods. Spacing between waveguides at $r=0 ~\mu\textrm{m}$ was set to $d=32\,\mu\textrm{m}$, and to achieve more uniform coupling between elliptic waveguides, their longer axes were oriented along the diagonal of the array (see schematics in Fig.~\ref{fig4}a and microphotographs of inscribed structure at different distances in Fig.~\ref{fig4}b). As one can see, such structure realizes the photonic Floquet higher-order insulator periodically switching between ``instantaneous'' topological and non-topological phases.
	
	\begin{figure}[t]
		\centering
		\includegraphics[width=\columnwidth]{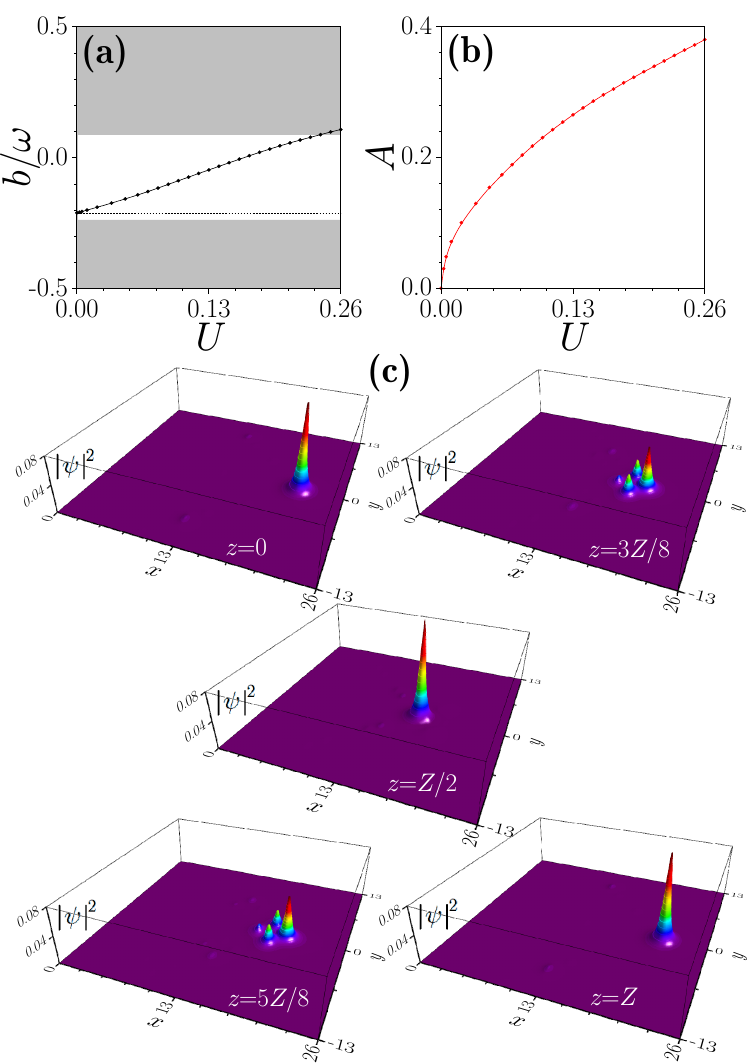}
		\caption{(Color online) Quasi-propagation constant (a) and $z$‐averaged peak amplitude (b) versus power $U$ for the $\pi$ solitons in 2D array. (c) Intensity distributions at different distances for a soliton with $U =0.2$. In all cases $r = 7\,\mu\textrm{m}$, $Z=49.5 ~\textrm{mm}$.}
		\label{fig5}
	\end{figure}
	
	Dependence of quasi-propagation constants $b$ of the Floquet eigenmodes of the 2D array, obtained from linear version of Eq.~(\ref{eq2}), on amplitude of waveguide oscillations $r$ shown in Fig.~\ref{fig4}c reveals the formation of 2D $\pi$ modes (red lines) that reside in the corners of the structure, but in comparison with 1D case they appear in sufficiently narrow range of oscillation amplitudes. This is a consequence of substantially more complex spectrum of static 2D SSH structures \cite{hu.light.10.164.2021} featuring four bands in topological phase (in contrast to only two bands in 1D SSH arrays), that in our case experience folding due to longitudinal array modulation, resulting in a very complex Floquet spectrum. For instance, quasi-propagation constants of 2D $\pi$ modes may overlap with the band, as it also happens with eigenvalues of usual corner modes in static higher-order insulators~\cite{hu.light.10.164.2021}. To obtain such spectrum, where topological gap can be opened by longitudinal array modulation, we had to select not too small depth of potential $p=5$ to ensure that the width of the bulk bands is comparable with the width of the longitudinal Brillouin zone and no too strong band folding occurs. Example of the $\pi$ mode performing periodic oscillations (only one corner is shown) is depicted in Fig.~\ref{fig4}d. The properties of 2D $\pi$ solitons, whose family was obtained from nonlinear Eq.~(\ref{eq2}) using iterative method, are summarized in Fig.~\ref{fig5}. As in the 1D case, quasi-propagation constant of 2D solitons crosses the gap with increase of power $U$ and enters into the band (Fig.~\ref{fig5}a). For the selected amplitude $r=7\,\mu\textrm{m}$ the soliton exists practically in the entire gap, because $b$ of linear $\pi$ mode from which it bifurcates is located near the lower gap edge (we have checked that this linear $\pi$ mode indeed falls into forbidden gap of bulk system by calculating quasienergy spectrum of periodic, i.e. infinite in the transverse direction, Floquet array). The average amplitude $A$ increases with $U$ (Fig.~\ref{fig5}b). The intensity distributions at different distances illustrating periodic $\pi$ soliton evolution in $z$ are presented in Fig.~\ref{fig5}c. Despite the fact that this state is 2D and oscillates strongly, the collapse is suppressed and one observes very robust propagation for all powers, when soliton resides in the gap. This conclusion was also supported by the results of propagation of weakly perturbed 2D $\pi$ solitons over large distances. For $r=7\,\mu\textrm{m}$ all such states in the gap were found stable.
	
	\begin{figure}[htpb]
	\centering
	\includegraphics[width=\columnwidth]{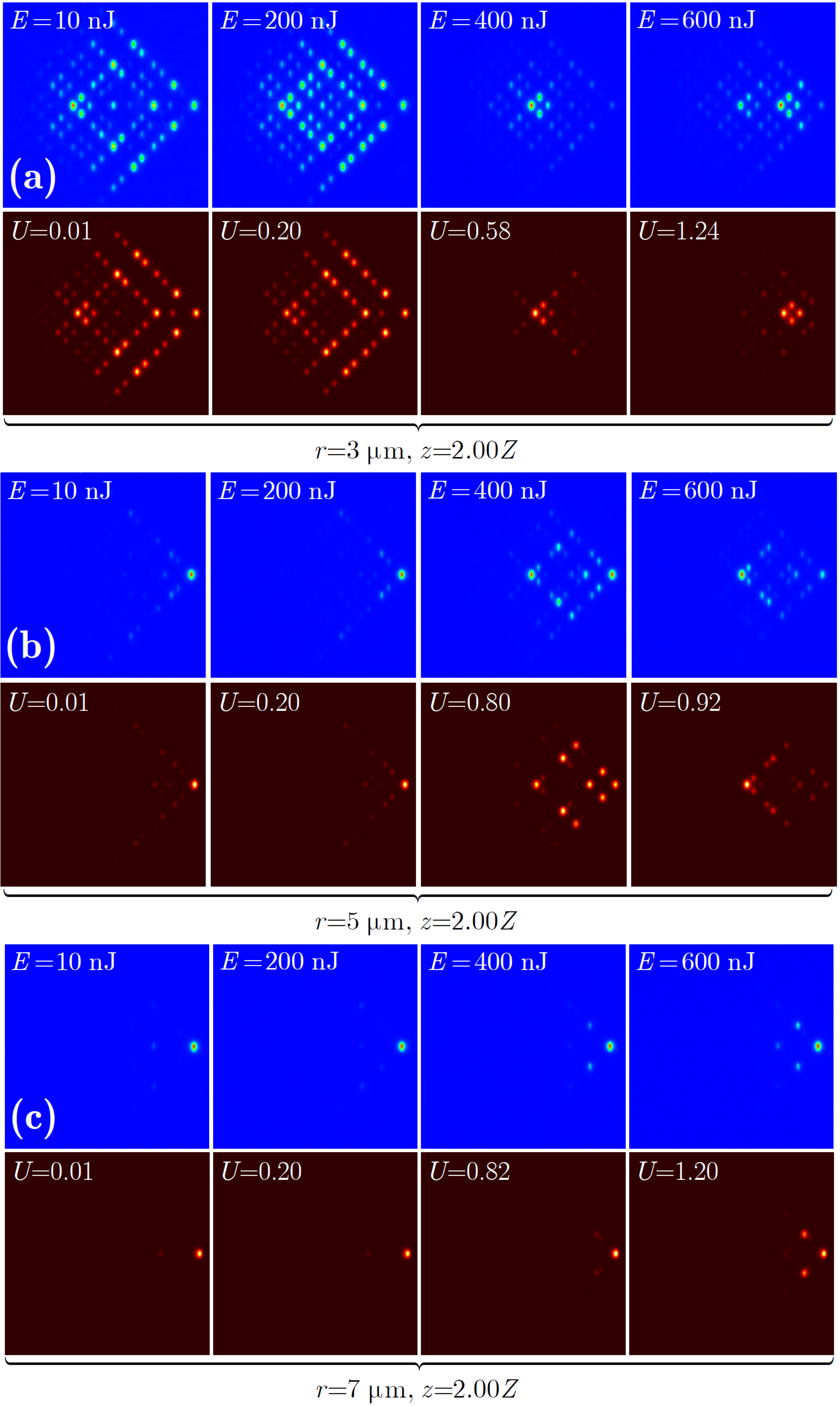}
	\caption{(Color online) Excitation of the $\pi$ solitons in 2D oscillating waveguide arrays for different amplitudes of the waveguide oscillations $r$ at $z= 2Z$. Top rows (blue background) show experimentally measured intensity distributions for different pulse energies $E$, while bottom rows (black background) show theoretically calculated output patterns for different powers $U$.}
	\label{fig6}
	\end{figure}
	
	To observe 2D $\pi$ solitons we inscribed the series of 2D oscillating arrays with various amplitudes of waveguide oscillations up to $r=9~\mu\textrm{m}$. Excitations in the right corner of the array were used, but it should be stressed that excitations of other corners yields nearly identical results due to high symmetry and uniformity of the array. At low pulse energies $E=10\,\textrm{nJ}$ one observes strong diffraction into the bulk at $r=3\,\mu\textrm{m}$ (Fig.~\ref{fig6}a), while efficient excitation of linear $\pi$ modes takes place at amplitudes $r\ge 5\,\mu\textrm{m}$ (Fig.~\ref{fig6}b and c). To the best of our knowledge, this constitutes the first observation of 2D $\pi$ modes in photonics. Increasing pulse energy at low $r\sim 3\,\mu\textrm{m}$ results first in concentration of light in the bulk of the sample and then its gradual displacement toward the corner (Fig.~\ref{fig6}a). By  contrast, for $r\ge 5\,\mu\textrm{m}$ one observes the formation of $\pi$ solitons, whose range of existence in terms of input power grows with increase of $r$. Thus, at $r=5\,\mu\textrm{m}$ the well-localized solitons form at pulse energies $E<300~\textrm{nJ}$, while at $E\sim 400~\textrm{nJ}$ strong radiation into the bulk occurs (Fig.~\ref{fig6}b) due to nonlinearity-induced shift into the allowed band. At $r=7\,\mu\textrm{m}$ one observes the formation of the $\pi$ solitons even at $E\sim 600~\textrm{nJ}$ (Fig.~\ref{fig6}c) with tendency for slight increase of secondary intensity maxima in soliton profile at highest power levels that is observed also in exact soliton solution of Eq.~(\ref{eq2}). Excitations in other corners of the array (e.g., top one) yield similar results confirming the $\pi$ soliton formation, while excitations in the bulk strongly diffract at these pulse energies.

	\section{Conclusion}
	
	We presented experimental observation of a new type of $\pi$ solitons in nonlinear Floquet system, where nontrivial topology arises from periodic modulation of the underlying photonic structure in evolution variable (along the light propagation path). Such solitons exist both in 1D and 2D geometries and they show exceptionally robust evolution due to practically absent radiative losses at considered periods and amplitudes of oscillations. The results obtained here may be used in the design of a class of topological Floquet lasers based on $\pi$ modes, for the control and enhancement of parametric processes, such as generation of new harmonics assisted by topology of the Floquet system, and for design of new types of on-chip all-optically controlled topological devices.

	\section*{Conflict of interest}
	The authors declare that they have no conflict of interest.
		
	\section*{Acknowledgments}
	This research is funded by the research project FFUU-2021-0003 of the Institute of Spectroscopy of the Russian Academy of Sciences and partially by the RSF grant 21-12-00096. Y. Z. acknowledges funding by the National Natural Science Foundation of China (Grant Nos.: 12074308). S. Z. acknowledges support by the Foundation for the Advancement of Theoretical Physics and Mathematics “BASIS” (Grant No.: 22-2-2-26-1).

	\section*{Author contributions}
	Yiqi Zhang and Yaroslav V. Kartashov formulated the problem. Sergei A. Zhuravitskii, Nikolay N. Skryabin, Ivan V. Dyakonov, and Alexander A. Kalinkin fabricated the samples. Antonina A. Arkhipova, Victor O. Kompanets, and Sergey V. Chekalin performed experiments. Yiqi Zhang performed numerical modeling. Yaroslav V. Kartashov, Sergei P. Kulik, and Victor N. Zadkov supervised the work. All co-authors took part in discussion of results and writing of manuscript.
	
	%	\bibliography{my_library}

\end{document}